\documentclass[reprint,aps,twocolumn,superscriptaddress,showpacs,
floatfix,preprintnumbers,amsmath]{revtex4-1}

\usepackage{graphicx}
\usepackage[usenames,dvipsnames]{xcolor}
\usepackage{ifpdf}
\usepackage{array,booktabs}
\usepackage{notoccite}
\usepackage{changes}
\usepackage{bm}

\ifpdf
  \usepackage[pdftex]{graphicx}
  \DeclareGraphicsExtensions{.pdf}
  \usepackage[dvipdfm,hyperindex=true]{hyperref}
\else
  \usepackage{graphicx}
  \DeclareGraphicsExtensions{.ps,.eps}
  \usepackage[dvipdfm,hyperindex=true]{hyperref}
\fi

\definecolor{linkcol}{rgb}{0.2,0.2,0.6}

\hypersetup
{
bookmarksopen=true,
pdfmenubar=true, 
pdfhighlight=/O, 
colorlinks=true, 
pdfpagemode=None, 
pdffitwindow=true, 
linkcolor=linkcol, 
citecolor=linkcol, 
urlcolor=linkcol 
}

\def\TN{\ensuremath{T_{\rm N}}}

\def\BaTi{Ba(TiO)Cu$_4$(PO$_4$)$_4$}

\def\Qb{\ensuremath{\mathbf{Q}}}
\def\Gb{\ensuremath{\mathbf{G}}}
\def\rb{\ensuremath{\mathbf{r}}}
\def\kk{\ensuremath{\mathbf{k}}}
\def\dr{\ensuremath{\rm d}}
\def\muB{\ensuremath{\mu_B}}

\begin{document}


\title{Magnetic structure of Ba(TiO)Cu$_4$(PO$_4$)$_4$ probed using spherical neutron polarimetry}

\author{P. Babkevich}
\email{peter.babkevich@gmail.com}
\affiliation{Laboratory for Quantum Magnetism, Institute of Physics, \'{E}cole Polytechnique F\'{e}d\'{e}rale de Lausanne (EPFL), CH-1015 Lausanne, Switzerland}
\author{L. Testa}
\affiliation{Laboratory for Quantum Magnetism, Institute of Physics, \'{E}cole Polytechnique F\'{e}d\'{e}rale de Lausanne (EPFL), CH-1015 Lausanne, Switzerland}
\author{K. Kimura}
\affiliation{Division of Materials Physics, Graduate School of Engineering Science, Osaka University, Toyonaka, Osaka 560-8531, Japan}
\author{T. Kimura}
\affiliation{Department of Advanced Materials Science, University of Tokyo, Kashiwa 277-8561, Japan}
\author{G. S. Tucker}
\affiliation{Laboratory for Quantum Magnetism, Institute of Physics, \'{E}cole Polytechnique F\'{e}d\'{e}rale de Lausanne (EPFL), CH-1015 Lausanne, Switzerland}
\affiliation{Laboratory for Neutron Scattering and Imaging, Paul Scherrer Institut, CH-5232 Villigen, Switzerland}
\author{B. Roessli}
\affiliation{Laboratory for Neutron Scattering and Imaging, Paul Scherrer Institut, CH-5232 Villigen, Switzerland}
\author{H. M. R\o nnow}
\affiliation{Laboratory for Quantum Magnetism, Institute of Physics, \'{E}cole Polytechnique F\'{e}d\'{e}rale de Lausanne (EPFL), CH-1015 Lausanne, Switzerland}

\begin{abstract}
The antiferromagnetic compound \BaTi\ contains square cupola of corner-sharing CuO$_4$ plaquettes, which were proposed to form effective quadrupolar order. To identify the magnetic structure, we have performed spherical neutron polarimetry measurements. Based on symmetry analysis and careful measurements we conclude that the orientation of the Cu$^{2+}$ spins form a non-collinear in-out structure with spins approximately perpendicular to the CuO$_4$ motif. Strong Dzyaloshinskii-Moriya interaction naturally lends itself to explain this phenomenon. The identification of the ground state magnetic structure should serve well for future theoretical and experimental studies into this and closely related compounds.
\end{abstract}

\maketitle

\begin{figure*}
\includegraphics[bb = 0 0 650 400,clip= ,width=0.6\textwidth]{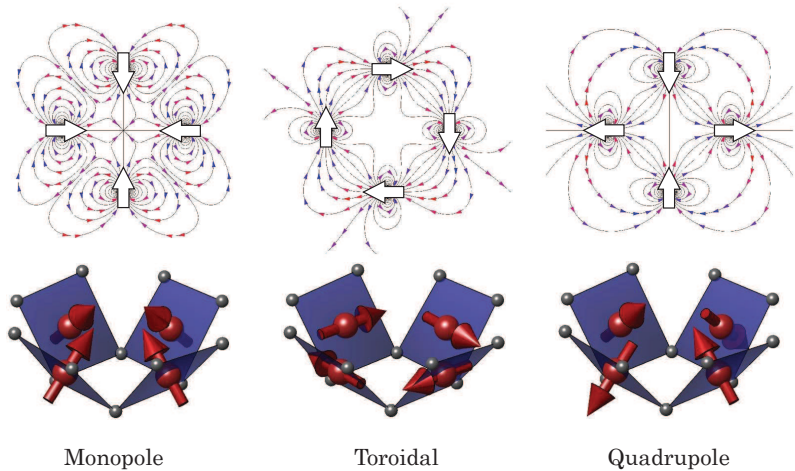}
\caption{Illustration of the arrangement of spins on the Cu$_{4}$O$_{12}$ plaquette to produce an effective monopolar, toroidal, or quadrupolar moment in \BaTi.}
\label{fig:0}
\end{figure*}

The magnetoelectric effect, that describes the coupling between magnetism and ferroelectricity, allows for the ability to control the material's magnetization using an electric field or polarization using a magnetic field; making it a promising avenue for the next generation of data storage materials. A linear magnetoelectric effect in magnetically ordered systems  necessitates the breaking of both the time reversal and the spatial inversion symmetry. The magnetic interaction energy of a magnetization density with an inhomogeneous magnetic field can be written as a multipole expansion containing monopole, toroid, and quadrupole moments, illustrated in Fig.~\ref{fig:0} \cite{spaldin-jpcm-2008}. All three moments change sign under time reversal or space inversion as necessary for the linear magnetoelectric effect. Although toroidal multipole moments have been shown to possess magnetoelectric activity, the recently discovered, \BaTi\ is believed to be the first experimental observation of magnetoelectric activity originating from magnetic quadrupole moments \cite{kimura-inorg-2016, kato-prl-2017}.

The analysis of our previous powder neutron diffraction measurements was able to identify two possible models for the magnetic structure of \BaTi\ \cite{kimura-natcomm-2016}; however, it was limited by: (i) weak magnetic Bragg reflections due to the small magnetic moment on Cu ions and (ii) the assumption of isotropic magnetic form factor for Cu systems, which is typically not the case. To ascertain the magnetic ordering in \BaTi\ we have performed polarized neutron scattering measurements. Spherical neutron polarimetry is a convenient, albeit rarely used tool for understanding complex magnetic structures. When applied, it can often provide unambiguous solutions to withstanding problems. In this article we describe spherical neutron polarimetry results to demonstrate that the magnetic structure of \BaTi\ can be uniquely identified. Furthermore, we present detailed methodology that could serve for future experiments aiming to utilize this technique.


\begin{figure*}
\includegraphics[clip= ,width=0.85\textwidth]{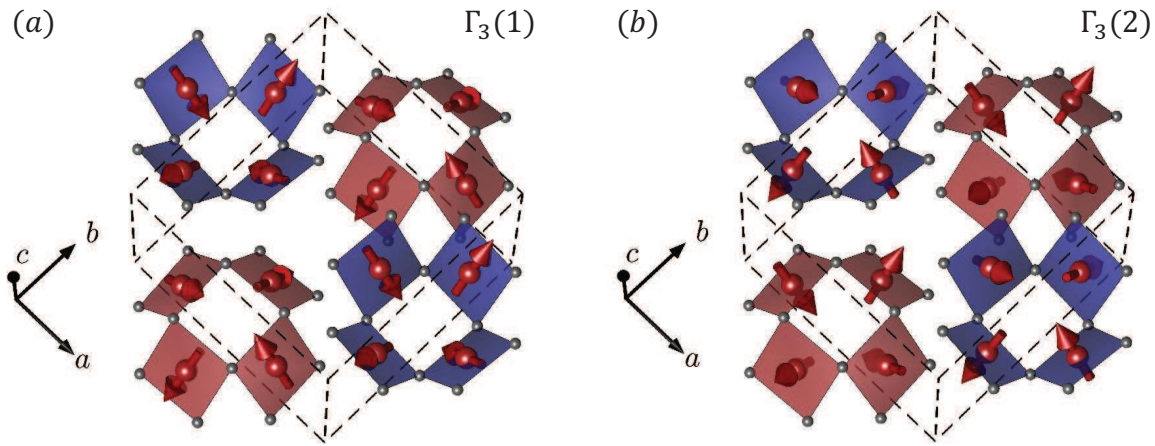}
\caption{Proposed magnetic structures described in the text of \BaTi\ obtained from spherical neutron polarimetry. (a) Moments are lying almost in the CuO$_4$ planes. (b) Moments are almost perpendicular to CuO$_4$ planes.}
\label{fig:1}
\end{figure*}

\section{Symmetry analysis}
\label{sec:sym_ana}

\begin{table*}
\centering
\begin{tabular}{c|c|c|c}
\hline
\hline
Element & Rotation matrix & IT notation  & Jones symbol  \\
 $g_n$  & {$M$}           &              &           \\
\hline
 $g_1$  & $\begin{pmatrix}
            1 & 0 & 0 \\
            0 & 1 & 0 \\
            0 & 0 & 1
          \end{pmatrix}$
          & 1
          & $(x,y,z)$ \\

 $g_2$  & $\begin{pmatrix}
            \bar{1} & 0 & 0 \\
            0 & \bar{1} & 0 \\
            0 & 0 & 1
          \end{pmatrix}$
          & 2\ $0,0,z$
          & $(-x,-y,z)$ \\

 $g_3$  & $\begin{pmatrix}
            0 & \bar{1} & 0 \\
            1 & 0 & 0 \\
            0 & 0 & 1
          \end{pmatrix}$
          & $4^+$\ $0,\frac{1}{2},z$
          & $(-y+1/2,x+1/2,z)$ \\

 $g_4$  & $\begin{pmatrix}
            \bar{1} & 0 & 0 \\
            0 & 1 & 0 \\
            0 & 0 & \bar{1}
          \end{pmatrix}$
          & $2(0,\frac{1}{2},0)$\ $\frac{1}{4},y,0$
          & $(-x+1/2,y+1/2,-z)$ \\

 $g_5$  & $\begin{pmatrix}
            0 & 1 & 0 \\
            \bar{1} & 0 & 0 \\
            0 & 0 & 1
          \end{pmatrix}$
          & $4^-$\ $\frac{1}{2},0,z$
          & $(y+1/2,-x+1/2,z)$ \\

 $g_6$  & $\begin{pmatrix}
            1 & 0 & 0 \\
            0 & \bar{1} & 0 \\
            0 & 0 & \bar{1}
          \end{pmatrix}$
          & $2(\frac{1}{2},0,0)$\ $x, \frac{1}{4} ,0$
          & $(x+1/2,-y+1/2,-z)$ \\

 $g_7$  & $\begin{pmatrix}
            0 & \bar{1} & 0 \\
            \bar{1} & 0 & 0 \\
            0 & 0 & \bar{1}
          \end{pmatrix}$
          & 2\ $x, \bar{x},0$
          & $(-y,-x,-z)$ \\

 $g_8$  & $\begin{pmatrix}
            0 & 1 & 0 \\
            1 & 0 & 0 \\
            0 & 0 & \bar{1}
          \end{pmatrix}$
          & 2\ $x, x,0$
          & $(y,x,-z)$ \\
\hline
\hline
\end{tabular}
\caption{Symmetry operators of space group $P42_12$ showing explicitly the rotational part $M$, IT notation as listed in the International Tables of Crystallography, and the Jones representation.}
\label{tab:sym_op}
\end{table*}

\begin{table*}
\centering
\begin{tabular}{ l  c c c c  c c c c c}
\hline
\hline
$\nu$
& $g_1$
& $g_2$
& $g_3$
& $g_4$
& $g_5$
& $g_6$
& $g_7$
& $g_8$
& MSG\\
\hline
1 &
  1  &  1  &  1        &     1     &        1      &       1     &        1       &      1
& $P42_12$\\
2 &
  1  &  1  &  1        &     1     &       $\bar{1}$      &      $\bar{1} $    &       $\bar{1}$       &     $\bar{1}$
& $P42_121'$\\
3 &
  1  &  1  & $\bar{1}$        &    $\bar{1}$     &        1      &       1     &       $\bar{1}$        &     $\bar{1}$
& $P42_121'$\\
4 &
  1  &  1  & $\bar{1}$         &    $\bar{1}$      &       $\bar{1}$       &      $\bar{1}$      &        1       &      1
& $P4'2_1'2$\\
5 &
  $\begin{pmatrix} 1 & 0  \\ 0 & 1 \end{pmatrix}$ &
  $\begin{pmatrix}\bar{1} & 0  \\ 0 & \bar{1} \end{pmatrix}$ &
  $\begin{pmatrix} 1 & 0  \\ 0 & \bar{1} \end{pmatrix}$ &
  $\begin{pmatrix}\bar{1} & 0  \\ 0 & 1 \end{pmatrix}$ &
  $\begin{pmatrix} 0 & 1  \\ 1 & 0 \end{pmatrix}$ &
  $\begin{pmatrix} 0 & \bar{1}  \\ \bar{1} & 0 \end{pmatrix}$ &
  $\begin{pmatrix} 0 & \bar{1}  \\  1 & 0 \end{pmatrix}$ &
  $\begin{pmatrix} 0 &  1  \\ \bar{1} & 0 \end{pmatrix}$ & -\\
\hline\hline
\end{tabular}
\caption{Character table of the little group $G_\kk$ showing how the IRs $\Gamma_\nu$ transform according to symmetry operations $g_1,\ldots,g_8$ in Table~\ref{tab:sym_op}. The final column gives the magnetic space group of each IR in the Belov-Neronova-Smirnova notation.
\label{tab:sym_elements}}
\end{table*}

\BaTi\ crystallizes in a chiral tetragonal structure with a space group $P42_12$ with lattice parameters of $a=9.56$\,\AA\ and $c=7.07$\,\AA\ \cite{kimura-inorg-2016}. The upward and downward square cupola of Cu$_4$O$_{12}$ are arranged in an alternating fashion in the tetragonal $ab$ plane, \cite{kimura-inorg-2016} shown in Fig.~\ref{fig:1}. In between the Cu$_4$O$_{12}$ cupola lies a non-magnetic layer composed of tetrahedra of PO$_4$ and pyramids of TiO$_5$. The crystallographic unit cell contains 8 Cu ions which are all equivalent to the general position $(0.27,0.99,0.40)$. The magnetic structure of \BaTi\ was previously studied in Ref.~\cite{kimura-natcomm-2016} using neutron powder diffraction. Antiferromagnetic order develops below $\TN = 9.5$\,K giving rise to magnetic reflections which can be indexed using a magnetic propagation wave vector $\kk = (0,0,0.5)$. Group representation theory can be used to identify the possible magnetic structures emanating from the paramagnetic group from which the magnetic order emerges. A number of software packages are available to perform such an analysis, such as Basireps \cite{fullprof}. Below we outline the steps used to calculate the possible magnetic structures.

\subsection{Representation analysis of magnetic structures}

The little group $G_\kk$ is a subset of symmetry elements within the paramagnetic space group $G_0$ ($P42_12$) which leave the propagation wavevector invariant under unitary transformation matrix $M$. In our case the little group contains all elements of $G_0$, which are listed in Table~\ref{tab:sym_op}. It is convenient to transform the representation of $G_\kk$ into irreducible representations (IRs) which are orthogonal to one another.

The magnetic representation $\Gamma_{\rm mag}$ is the result of the symmetry operations on the  position (polar) and spin (axial) vectors. The two are independent and can be treated separately. The former permutes the atomic positions $\mathbf{r}$ such that, $g_n \mathbf{r}_i = \mathbf{r}_j$. The magnetic spin $\mathbf{S}$ must obey axial vector property and remain invariant under an inversion, or $\mathbf{S}' = |M|M \mathbf{S}$. The magnetic representation is then a tensor product of the permutation and axial representations,
\begin{align}
  \Gamma_{\rm mag} & = \Gamma_{\rm perm} \times \Gamma_{\rm axial}, \\
  \chi_{\rm mag} & = \chi_{\rm perm} \chi_{\rm axial}.
\end{align}
The character $\chi$ of permutation and axial vector representations is simply given by the trace of the respective representations. Any magnetic representation is reducible to block-diagonal form by a summation over the IRs $\Gamma_\nu$. The magnetic representation can then be described as,
\begin{align}
\Gamma_{\rm mag} &= \sum_\nu n_\nu \Gamma_\nu, \label{eqn:Gamma_mag}\\
n_\nu &= \frac{1}{n(G_{\mathbf k})}\sum_{g \in G_{\mathbf k}}
\chi_{\rm mag}(g)\chi_{\Gamma_\nu}(g)^\ast.
\end{align}
The value of $n_\nu$ tells us how many distinct basis vector we can expect for each irreducible representation. In \BaTi, the magnetic representation at the Cu site with Wyckoff position $8g$ can be decomposed into a direct sum of irreducible representations as, $\Gamma_{\rm mag}(8g) = 3\Gamma_1 + 3\Gamma_2 + 3\Gamma_3 + 3\Gamma_4 + 6\Gamma^{(2)}_5$. All IRs are one-dimensional, except $\Gamma^{(2)}_5$, which is two-dimensional. The character table for $G_\kk$ is given in Table~\ref{tab:sym_elements}.

The basis vectors $\bm{\psi}$ are calculated using the projection operator technique by using a trial functions along crystallographic axes $\bm{m}_a = (1,0,0)$, $\bm{m}_b = (0,1,0)$ and $\bm{m}_c = (0,0,1)$. The projection operator formula to find the basis vector $\bm{\psi}$ for magnetic representation $\Gamma_\nu$ is given as,
\begin{equation}
\bm{\psi}_{\alpha\nu} = \sum_{g\in G_{\mathbf k}} \chi_\nu^{\ast}(g)
\sum_n \delta_{n,g_n}|M(g)|M(g)\bm{m}_\alpha
\end{equation}
where $\chi(g)$ is the character of the little group $G_\kk$. The spin distribution of the $j$th atom can be expressed as the Fourier transform of the linear combination of basis vectors, such that for a single propagation wavevector,
\begin{equation}
  \mathbf{S}_j = \sum_n C_n \bm{\psi}_n e^{-i\kk\cdot\mathbf{t}} + c.c.,
\end{equation}
where the coefficients $C_n$ can, in general, be complex. The Fourier coefficients obtained from the basis function calculated for the $\Gamma_3$ IR for the Cu sites resolved along crystallographic axes are,
%
%
\begin{align*}
&1.(u,v,w);&
&2.(\bar{u},\bar{v},w);&
&3.(v,\bar{u},\bar{w});&
&4.(u,\bar{v},w)\\
&5.(\bar{v},u,\bar{w});&
&6.(\bar{u},v,w);&
&7.(\bar{v},\bar{u},\bar{w});&
&8.(v,u,\bar{w})
\end{align*}
The parameters $u$, $v$ and $w$ are free parameters of the possible magnetic structure which are to be determined experimentally and the labels refer to the symmetry operators $g_1,\ldots,g_8$. Equivalent calculations can be made for the other IRs in $\Gamma_{\rm mag}$. Rietveld refinement of the neutron powder diffraction data showed that $\Gamma_3$ gives the best agreement
\footnote{The case of $\Gamma^{(2)}_5$ can produce a magnetic structure with an amplitude modulated moment, which seems unlikely.}.

Within the $\Gamma_3$ IR, we find two solutions which give similar quality of fit to the observed magnetic pattern. We label these models as $\Gamma_3(1)$ and $\Gamma_3(2)$ with Fourier coefficients $(u,v,w)=(0.50(1),0.36(2),0.58(2))$ and $(0.49(1),0.0(1),-0.62(2))$, respectively. The magnetic structures of the two models closely resembles that illustrated in Fig.~\ref{fig:1}. We note that both $\Gamma_3(1)$ and $\Gamma_3(2)$ contain the component of the magnetic quadrupole moment which is illustrated in Fig.~\ref{fig:0}. In the case of $\Gamma_3(1)$ the moments are confined approximately in the plane of the CuO$_4$. Conversely, in the $\Gamma_3(2)$ model, the moments are approximately perpendicular to the CuO$_4$ planes and form a \textit{two-in-two-out} type arrangement within a Cu$_4$O$_{12}$ plaquette. The goodness-of-fit to the powder neutron diffraction data was found to be slightly better for the $\Gamma_3(2)$ model. However, the refinement suffers from two problems. First, the magnetic form factor (see Section~\ref{sec:magn_int}) was assumed to be isotropic, which is typically not the case for Cu$^{2+}$ ions. Furthermore, due to covalency, the magnetic form factor can be strongly modified \cite{walters-natphys-2009}. Second, at larger $|\Qb|$, the magnetic signal is rather weak due to the small magnetic moment of around 0.8\,$\mu_B$ and magnetic and structural Bragg peak overlap makes fitting difficult. Therefore, we turn to polarized neutron scattering to try to confirm and refine the complex, non-collinear magnetic structure found from powder neutron diffraction in \BaTi.

\subsection{Domains}

In order to accurately model the polarization matrices that are measured in spherical neutron polarimetry, we must consider the possible domains that could exist when the symmetry of the ordered magnetic structure is lower than that of the paramagnetic phase \cite{chatterji-book}. In the present case, we find that $G_\kk$ contains all the symmetry elements of the paramagnetic space group. This means that translation symmetry is preserved on applying the symmetry operators and no configuration domains (or \kk-domains) are produced.

An interesting property of \BaTi\ is the chiral crystal structure in which there is no roto-inversion axis. In this case, a pair of enantimorphs can be formed that are related by a spatial inversion. A polarized light beam will be rotated when traversing through such a sample with the rotation being sensitive to the structural chirality. Indeed, such measurements have demonstrated the presence of structural chiral domains in \BaTi\ \cite{kimura-inorg-2016}.

\section{Magnetic cross-section}
\label{sec:magn_int}

In order to derive the polarization matrices, we present a brief account of the neutron scattering theory behind it \cite{squires-book,shirane-book,roessli-book-2001}. The partial differential scattering cross-section in an elastic neutron scattering measurement can be described by,
\begin{equation}\label{eq:diff}
\frac{\dr \sigma}{{\dr} \Omega}  =
\sum_{i,f} P(\lambda_i)
| \langle  \lambda_f , \sigma_f
| \sum_j e^{i \Qb\cdot\rb_j} U_j|
\lambda_i, \sigma_i \rangle|^2
\delta(E).
\end{equation}
This gives the probability that a neutron is scattered into a solid angle $\Omega$ without transferring any energy to the system. The initial (final) states of the neutron and sample are given by $\sigma_i$ and $\lambda_i$ ($\sigma_f$ and $\lambda_f$), respectively. The statistical weight factor for an initial state $|\lambda_i\rangle$ is given by $P(\lambda_i)$. The last term ensures energy conservation during the scattering process, which in the present case will be restricted to purely elastic scattering. The atomic scattering amplitude for the $j$th atom at position $\rb_j$ is given as,
\begin{equation}\label{eq:atom_ampl}
U_j =  b^{\rm coh}_j + \frac{b_j^{\rm inc}}{\sqrt{I(I+1)}}\mathbf{I}_j \cdot \bm{\sigma} - p_j \mathbf{S}_{\perp j}\cdot\bm{\sigma}.
\end{equation}
The scattering length operator for the interaction between neutrons and nuclei consists of both coherent and incoherent scattering lengths $b$ which can contribute to the total scattering cross-section. The Pauli spin-operator $\bm{\sigma}$ is the normalized neutron spin operator and the nucleus spin operator is $\mathbf{I}$.

The last term in Eq.~(\ref{eq:atom_ampl}) describes the magnetic scattering of the neutrons by the sample. The factor $p = (\gamma r_0)f(\Qb)/2$ for a spin-only moment, where the gyromagnetic ratio $\gamma = 1.913$ and the classical electron radius $r_0 = 2.82$\,fm. The magnetic form factor $f(\Qb)$ corresponds to the Fourier transform of the unpaired spin density on an atom. The magnetic interaction vector, $\mathbf{S}_\perp = \hat{\Qb}\times \mathbf{S} \times \hat{\Qb}$ expresses the fact that only magnetization perpendicular to \Qb\ can scatter neutrons. In the case of coherent magnetic scattering of unpolarized neutrons from a magnetically ordered crystal, the elastic differential scattering cross-section is derived from Eqs.~(\ref{eq:diff}) and (\ref{eq:atom_ampl}) as,
\begin{equation}
  \frac{\dr \sigma}{\dr \Omega} \propto |\mathbf{F}(\Qb)|^2\delta(\Qb + \mathbf{G} \pm \mathbf{k})
\end{equation}
for a given propagation wavevector $\kk$ and reciprocal lattice wavevector $\mathbf{G}$. The structure factor is found as,
\begin{equation}\label{eq:FM}
  \mathbf{F}(\Qb) = \sum_j p_j \langle  \mathbf{S}_{\perp j}  \rangle e^{i\Qb\cdot\rb_j}e^{-W_j},
\end{equation}
which includes the Debye-Waller factor $e^{-W_j}$ and $\langle \mathbf{S}_{\perp} \rangle$ contains the thermally averaged expectation value of the spin perpendicular to \Qb.

\subsection{Polarized neutron scattering}
Polarized neutron scattering makes use of the incident and outgoing neutron spin state to provide additional information about the magnetic system. The polarization of a neutron beam is a statistical quantity defined as the expectation value of an ensemble of neutron spins. The scattering of neutron from a sample can in general reorient the neutron moment from one orientation to any other in three-dimensions. This process can be neatly described by a polarization matrix $P_{\alpha\beta}$, which consists of measuring 18 different scattering intensities in the spin-flip and non-spin-flip channels, $\sigma(\alpha,\beta)$ and $\sigma(\alpha,-\beta)$, respectively. The initial spin direction is defined by $\alpha$ and the final direction by $\beta$, such that,
\begin{equation}\label{eq:Pab}
  P(\alpha,\beta) = \frac{\sigma(\alpha,\beta) - \sigma(\alpha,-\beta)}{\sigma(\alpha,\beta) + \sigma(\alpha,-\beta)}.
\end{equation}
In neutron polarimetry, it is useful to define $x$ as parallel to \Qb, $z$ perpendicular to the scattering plane, and $y$ completes that the right-handed coordinate system. The $\alpha$ and $\beta$ are defined in this coordinate system,
\begin{align*}\label{eq:alpha_beta}
    | x \rangle &= \frac{1}{\sqrt{2}} \begin{pmatrix} 1 \\ 1 \end{pmatrix}, \quad
  & | y \rangle &= \frac{1}{\sqrt{2}} \begin{pmatrix} 1 \\ i \end{pmatrix}, \quad
  & | z \rangle &= \begin{pmatrix} 1 \\ 0 \end{pmatrix} \\\nonumber
    | \bar{x} \rangle &= \frac{1}{\sqrt{2}} \begin{pmatrix} 1 \\ \bar{1} \end{pmatrix}, \quad
  & | \bar{y} \rangle &= \frac{1}{\sqrt{2}} \begin{pmatrix} 1 \\ \bar{i} \end{pmatrix}, \quad
  & | \bar{z} \rangle &= \begin{pmatrix} 0 \\ 1 \end{pmatrix} \nonumber
\end{align*}

From Eqs.~(\ref{eq:diff}) and (\ref{eq:atom_ampl}) we will in general obtain nuclear, magnetic or a mix of nuclear-magnetic interference terms upon squaring. However, for present case of \BaTi, the magnetic propagation wavevector is $(0,0,0.5)$ and therefore the magnetic and nuclear reflections will occur at different positions in reciprocal space.

The neutron and electron (which is responsible for the magnetization) coordinates are independent, which allows us to separate the matrix element in Eq.~(\ref{eq:diff}) to $\langle  \sigma_{\rm f} | \bm{\sigma} | \sigma_{\rm i} \rangle \langle  \lambda_{\rm f} | \sum_j e^{i\Qb\cdot\rb_j} p_j \mathbf{S}_\perp | \lambda_{\rm i} \rangle$, where the second term is equivalent to $\mathbf{F}(\Qb)$ in Eq.~(\ref{eq:FM}). We define, $\mathbf{F}(\Qb) = (0,F_y,F_z)$ -- resolved along our chosen coordinate system $\{x,y,z\}$. Due to the cross-product in $\mathbf{S}_\perp$, the contribution of magnetization along $x$ is zero. Therefore, we can define the scattering cross-sections in Eq.~(\ref{eq:Pab}) as,
\begin{equation}\label{eq:xsec}
  \sigma(\alpha,\beta) \propto |\langle  \beta |
  \begin{pmatrix}  F_z & -i F_y \\ i F_y & -F_z \end{pmatrix}| \alpha \rangle|^2.
\end{equation}
Since the matrix is measured at a particular \Qb\ point and consists of normalized intensity in Eq.~(\ref{eq:Pab}), the polarization analysis is not sensitive to the magnetic form factor contained in $p_j$, nor to the size of the magnetic moment. In practice, measuring magnetic Bragg peaks at larger $|\Qb|$ becomes increasingly challenging as the magnetic form factor decreases the scattering intensity. For the special case of longitudinal neutron polarimetry at $\Qb = \Gb + \kk$,
\begin{align}
  \sigma(x,x) &= 0,        &\sigma(x,\bar{x}) &= |\mathbf{F}|^2 + i (\mathbf{F} \times \mathbf{F}^\ast) \nonumber\\
  \sigma(y,y) &= |F_y|^2,  &\sigma(y,\bar{y}) &= |F_z|^2 \nonumber\\
  \sigma(z,z) &= |F_z|^2,  &\sigma(z,\bar{z}) &= |F_y|^2 \nonumber
\end{align}
We note that the $\sigma(x,\bar{x})$ is sensitive to a chiral magnetic structure through the term $i (\mathbf{F} \times \mathbf{F}^\ast)$. In the case of perfect beam polarization and a single-domain structure, the polarization matrix can be found as,
\begin{align}
P &=
\begin{pmatrix}
\!-1 & 0 & 0\\
\ C & \!\!\!\!-A & B\\
\ C & B & A\\
\end{pmatrix}
\label{eqn:polmat_ABCD}\\
AD &= |F_z|^2 - |F_y|^2,\nonumber\\
BD &= F_y F_z^\ast + F_y^\ast F_z,\nonumber\\
CD &= i(F_y F_z^\ast - F_y^\ast F_z),\nonumber\\
D  &= |F_y|^2 + |F_z|^2.
\label{eqn:ABCD}
\end{align}
The $P(x,x)$ element readily identifies the nature of the reflection: $P(x,x) = +1$ for a nuclear Bragg peak and $P(x,x) = -1$ for a magnetic one. The coherent scattering from the nuclear structure does not contain any spin-dependence in Eq.~(\ref{eq:atom_ampl}), which means that the initial spin-state of the neutron will be preserved after scattering from the nuclear structure. This is an important consequence for polarized neutron scattering which allows us to cleanly separate the signal originating from coherent nuclear or magnetic scattering processes. It should be noted that the nuclear spin incoherent scattering can flip the spin and thereby contribute to a featureless spin-flip background.

The chiral term in Eq.~(\ref{eqn:ABCD}) can be equivalently expressed as $CD = i(\mathbf{F} \times \mathbf{F}^\ast)$ and is a signature of non-collinear magnetic order. For a single-domain, the summation $A^2 + B^2 + C^2 =1$ will hold for ideal beam polarization. Magnetic domains can depolarize the neutron beam such that $A^2 + B^2 + C^2 \leq1$. Symmetry consideration are necessary to account for this.

In the 1960s, Blume and Maleyev established equations that were useful in gaining insight into how the different scattering processes affect the different elements of the polarization matrix \cite{blume-pr-1963, izyumov-sovphys-1962}. However, this formulation was restricted to the single-domain case only, which does not hold true for most magnetic systems. Herein we shall outline the methodology for treating multi-domain structures. In the case of multi-domain sample, the scattering cross-section in Eq.~(\ref{eq:Pab}) becomes,
\begin{equation}
  \sigma(\alpha,\beta) \rightarrow \sum_n f_n \sigma_n(\alpha,\beta),
\end{equation}
where the fraction of the $n$th domain is given by $f_n$.

\subsection{Imperfect beam polarization correction}

In practice, 100\% neutron spin polarization is not possible. The incident and scattered beam will have a polarization efficiency of $0 < \eta_i, \eta_f < 1$. This will cause some neutrons to scatter into the wrong channel that needs to be corrected when comparing measured matrices with the calculated ones. For a measured scattering cross-section $\sigma_{\rm m}(\alpha,\beta)$, we must consider an ensemble average of $\eta_i | \alpha \rangle$ neutrons with the correct polarization and $(1 - \eta_i) | -\alpha \rangle$ with the wrong polarization. Considering similarly the out-going neutron beam polarization, gives the corrected scattering cross-section as,
\begin{align}\label{eq:pol_corr}
\sigma_{\rm c}(\alpha,\beta) = &\eta_i \eta_f \sigma(\alpha,-\beta)\nonumber\\
																&+ \eta_i (1 - \eta_f) \sigma(\alpha,-\beta)\nonumber\\
																&+ (1-\eta_i) \eta_f \sigma(-\alpha,\beta) \nonumber\\
																&+ (1-\eta_i) (1-\eta_f) \sigma(-\alpha,-\beta).
\end{align}
Similarly by setting $\alpha \rightarrow -\alpha$ and/or $\beta \rightarrow -\beta$ one can obtain the scattering cross-section for other channels. A good estimate of the polarization efficiencies can be obtained by measuring a Bragg reflection which is either purely nuclear or magnetic in origin. It is useful to define the spin-flip ratio, $R$, which for a nuclear reflection and $\eta_i = \eta_f = \eta$ is,
\begin{equation}
   R = \left(\frac{\sigma_{\rm NSF}}{\sigma_{\rm SF}}\right)_{\rm m} =
   \frac{1}{2 \eta(1-\eta)} - 1,
\end{equation}
by measuring the ratio of the intensities in the spin-flip (SF) and non-spin-flip (NSF) channels. A flipping ratio of 12 will correspond to a neutron beam polarization efficiency of 96\%. However, some caution needs to be taken particularly when working on systems with short-range magnetic order with a focusing monochromator and/or analyzers as $R$ for nuclear resolution-limited reflections can be somewhat higher than diffuse magnetic scattering owing to spatial distribution of neutron beam polarization.

\section{Experimental results}
\label{sec:results}

%
%

%

\begin{table}
\centering
\begin{tabular}{ l | r | r}
\hline
\hline
SNP+ND(WISH)     &   $\Gamma_3(1)$               & $\Gamma_3(2)$ \\
\hline
$u$ (\muB)            &   0.56(2)                & 0.56(2) \\
$v$ (\muB)            &   $-$0.01(1)             & 0.03(1) \\
$w$ (\muB)            &   0.59(2)                & $-$0.57(2) \\
$m_0$ (\muB)          &   0.81(1)                & 0.80(1) \\
levo-domain           &   64(7)\%                & 36(7)\%\\
$\chi^2_\nu$          &   18.9                   & 18.9 \\
\hline
\hline
\end{tabular}
\caption{The Cu spin direction has been obtained by fitting the spherical neutron polarimetry data. The moment direction $(u,v,w)$ are normalised to the moment size $m_0$ obtained from fitting WISH powder data. The refined levo chiral domain population is also shown. The values in parentheses indicate 1 standard-deviation uncertainties in the fit parameters.
\label{tab:mag_str}}
\end{table}

\begin{table}
\begin{tabular}{ l | r | r}
\hline
\hline
                &   $\Gamma_3(1)$               & $\Gamma_3(2)$ \\
\hline
ND(D20)         &   16.4\%                      & 11.4\% \\
SNP+ND(D20)     &   32.6\%                      & 18.8\% \\
ND(WISH)        &   18.5\%                      & 11.5\% \\
SNP+ND(WISH)    &   21.3\%                      & 19.4\% \\
SNP+ND(WISH) (4,5)      &   (35.5\%, 42.9\%)    & (18.0\%, 14.3\%) \\
\hline
\hline
\end{tabular}
\caption{Results of magnetic structure refinement based on neutron powder diffraction (ND) measurements obtained from the WISH and D20 diffractometers and single-crystal spherical neutron polarimetry (SNP) for the two proposed magnetic structures. In the case of SNP+ND, the magnetic structure from SNP was used but the moment size was refined using ND data. The SNP+ND(WISH) (4,5) corresponds to the goodness-of-fit for the diffraction patterns obtained from detector banks covering the smallest $|\Qb|$ range.
\label{tab:nd}}
\end{table}

\subsection{Experimental setup}
Spherical neutron polarimetry measurements were performed using MuPAD configuration of the TASP spectrometer configuration at SINQ \cite{fischer-physicab-1997,semadeni-physicab-2001,janoschek-physicab-2007}. Incident neutrons of wavelength of 1.97\,\AA$^{-1}$ were used for the measurements. A single-crystal \BaTi\ sample of 0.6\,g grown by the flux method \cite{kimura-inorg-2016} was mounted on an Al holder. Measurements of the flipping ratio were performed on the $(200)$, $(220)$, and $(002)$ nuclear reflections giving $R=13.2$ corresponding to $\eta_i = \eta_f = 96.4$\%. The modeled polarization matrices take this non-ideal beam polarization into account as defined in Eq.~(\ref{eq:pol_corr}). Two sample orientation geometries were used to access the $(h0l)$ and $(hhl)$ scattering planes. Complete normal $P(\alpha,\beta)$ and negative $P(-\alpha,\beta)$ polarization matrices were recorded at 1.5 and 20\,K to eliminate the contributions from the systematic errors and background. In total 26 polarization matrices were used in the analysis, with around 1 hour counting time per matrix.

\subsection{Polarization matrix simulations}

\begin{figure*}
\includegraphics[clip= ,width=0.7\textwidth]{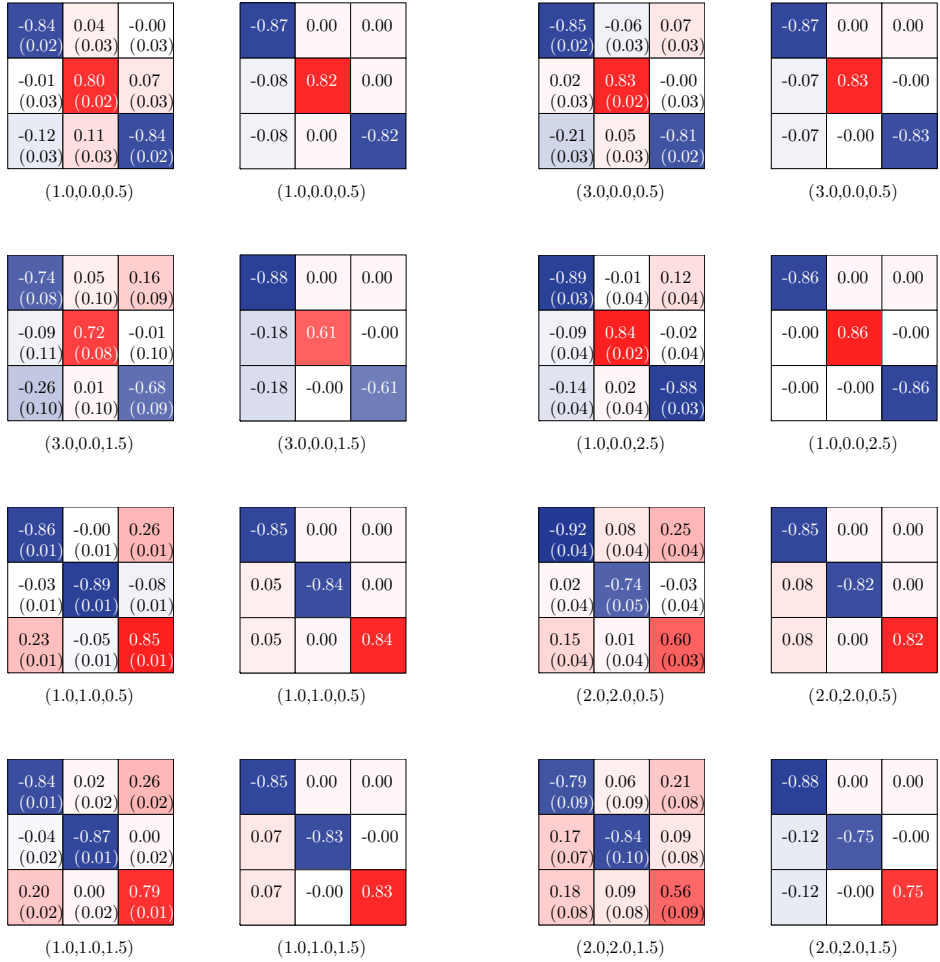}
\caption{A selection of the polarization matrices recorded for different $(hkl)$-positions indicated below the polarization matrices. Blue, white, and red colors represent $P(\alpha,\beta) = -1$, 0, and +1, respectively. Uncertainty in the matrix element is given in parenthesis. A background collected at 20\,K has been removed from the data measured at 1.5\,K. Matrices without parenthesis indicate simulation matrices. The simulated polarization matrices were calculated for the $\Gamma_3(2)$ magnetic structure given in Table~\ref{tab:mag_str}. A correction for beam polarization efficiency of 96.4\% has been applied to the calculated polarization matrices.}
\label{fig:2}
\end{figure*}

Figure~\ref{fig:2} illustrates some of the polarization matrices recorded.  We note that in our data $P(y,x) \neq P(z,x)$ and for some reflections $P(x,z)\neq 0$. These are probably caused by small gaps in the mu-metal shielding which result in systematic errors. To mitigate these errors, we have collected equivalent reflections and measured negative polarization matrices. An alternate scenario could be that there is a small amount of nuclear-magnetic interference, perhaps originating from some sort of a superstructure; however, our data is insufficient to provide further insight.

No magnetic intensity was found for reflections $(0,0,0.5)+(0,0,l)$ for $l = 0,1,2$. One might naively expect this to reflect that spins are all parallel to the $c$-axis. However, this can be easily shown not to hold true. If we examine the $P(y,y)$ and $P(z,z)$ elements at $(1,1,0.5)$ for example, we find measured values of $-0.89(1)$ and $+0.85(1)$, respectively [see Fig.~\ref{fig:2}]. Assuming a beam polarization efficiency of 96.4\% and spins along the $c$ axis results in $P(y,y) = +0.86$ and $P(z,z) = -0.86$, i.e., exactly opposite to what we observe experimentally. Indeed, a very poor goodness-of-fit to the complete data set of $\chi^2_\nu = 2000$ is found for such a model.

A key advantage of symmetry analysis is that it greatly reduces the number of free parameters. In the case of \BaTi\ we are left, for each possible IR, with a refinement of 3 parameters: the polar and zenith angles of the Cu spin and the levo-dextro domain population. We treat levo as the domain in which Cu ion is situated at $(0.27,0.99,0.40)$ and dextro as the structural chiral domain related by a spatial inversion. Fitting the complete set of the polarization matrices that have been collected, we find two possible solutions, shown in Table~\ref{tab:nd}, with identical quality of fit of $\chi^2_\nu = 18.9$
\footnote{We note that the $\chi^2_\nu$ only includes the random errors and not the systematic ones, resulting in values somewhat larger than 1.}.
The measured and simulated polarization matrices for the $\Gamma_3(2)$ structure are shown in Fig.~\ref{fig:2}. These closely match the solutions obtained from neutron powder diffraction \cite{kimura-natcomm-2016}. The two solutions are shown in Fig.~\ref{fig:1}. $\Gamma_3(1)$ is characterized by spins lying in the CuO$_4$ plane (within about 5$^\circ$), while in $\Gamma_3(2)$ they are nearly perpendicular to the plane, being approximately 5$^\circ$ from the normal of the CuO$_4$ plane. It is interesting to note that $\Gamma_3(1)$ can be mapped into $\Gamma_3(2)$ spin structures, and vice versa, through a displacement of the moments by $(0.5,0.5,0.2)$. If the $z$ component were zero, the magnetic structure factors of the two models would be the same; however, this small shift along the $c$ axis between the Cu$_4$O$_{12}$ plaquettes gives a discernable, albeit small, difference between the two models.

To elucidate the magnetic structure of \BaTi, we return to neutron powder diffraction measurements. Two experiments have been carried out so far on the same \BaTi\ powder sample using D20 at ILL and WISH at ISIS diffractometers whose results are reported elsewhere \cite{kimura-unpub,kimura-natcomm-2016}. In Table~\ref{tab:nd} we present the goodness-of-fit values obtained by either fitting the data allowing the moment size and direction to vary or fixing the moment direction as obtained from spherical neutron polarimetry. We find that in each case there is a better fit obtained for the $\Gamma_3(2)$ spin structure. The calculations were performed assuming an isotropic magnetic form factor. However, the result is almost exactly the same as when assuming Cu with $d_{x^2-y^2}$ orbitals along the Cu-O bonds. The difference is small because (i) the magnetic reflections that we have collected are at relatively small $|\Qb|$ such that the anisotropic and isotropic magnetic form factors are similar and (ii) the CuO$_4$ cupolas are tilted out of the $ab$ plane such that the net electronic spin density from each cupola would be expected to be relatively uniform.

\begin{figure}
\includegraphics[clip= ,width=0.95\columnwidth]{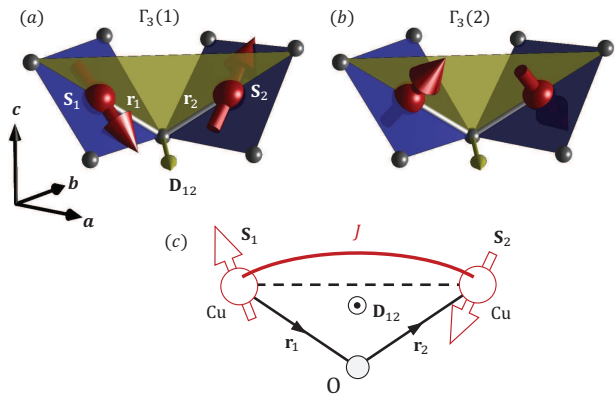}
\caption{Illustration of the DM interaction at play in \BaTi. Panels (a) and (b) show the nearest-neighbor Cu spins, each connected to 4 O atoms in case of $\Gamma_3(1)$ and $\Gamma_3(2)$ spin structures. The yellow plane represents the normal to the Cu-O-Cu bond, with the direction of the DM vector shown in yellow. (c) Simplified diagram to illustrate the DM interaction.}
\label{fig:3}
\end{figure}

Let us consider just the nearest-neighbor Cu spins, as depicted in Fig.~\ref{fig:3}. We note that the dominant exchange path between Cu ions is likely to be through the shared O atom. The large displacement of the O from the line connecting Cu-Cu sites [see Fig.~\ref{fig:3}(c)] implies first that the superexchange interaction $J$ is likely to be small; and second that the Dzyaloshinskii-Moriya (DM) interaction could be strong \cite{dzyaloshinsky-jopcs-1958, moriya-pr-1960}. The antiferromagnetic exchange interaction encourages the spins to be antiparallel while the DM interaction would favour a non-collinear spin arrangement. For two spins, the DM interaction has the form,
\begin{equation}\label{eq:DM}
  \mathcal{H}_{\rm DM} = -\mathbf{D}_{12}\cdot (\mathbf{S}_1\times\mathbf{S}_2),
\end{equation}
where the DM vector $\mathbf{D}_{12}\propto \lambda \mathbf{r}_1\times\mathbf{r}_2$ and $\lambda$ is the spin-orbit coupling. The vector connecting Cu and O is given by $\mathbf{r}$. In the case of \BaTi, we would expect the DM vector to be mostly in the $ab$-plane as shown in Figs.~\ref{fig:3}(a) and \ref{fig:3}(b). The most energetically favourable spin configuration in the presence of strong DM interaction would be where the spins lie in the plane normal to $\mathbf{D}_{12}$. This scenario is realized for the case of the $\Gamma_3(2)$ model where the spins are found to be almost normal to $\mathbf{D}_{12}$, see Fig.~\ref{fig:3}(b). For $\Gamma_3(1)$, the $\mathbf{S}_1\times\mathbf{S}_2$ vector is approximately 130$^\circ$ from $\mathbf{D}_{12}$, while for $\Gamma_3(2)$ this is just 20$^\circ$. Therefore, we would naively expect the DM interaction to stabilize the $\Gamma_3(2)$ rather than the $\Gamma_3(1)$ spin structure. A large DM interaction contribution has been proposed theoretically for \BaTi\ to reproduce the bulk magnetization measurements \cite{kato-prl-2017}. Moreover, inelastic neutron scattering measurements on \BaTi\ show a large spin-gap relative to the total bandwidth of the magnetic excitations \cite{kimura-natcomm-2016}; these results would be consistent with the present scenario where a strong DM interaction is responsible for the anisotropy and large spin-gap and demonstrates that our results are consistent.

\section{Conclusions}

Spherical neutron polarimetry is a powerful technique for probing systems where structural and magnetic signals are intertwined and/or magnetic structure is non-collinear. While standard methods of single-crystal and powder diffraction are able to shed light on magnetic ordering of systems, neutron polarimetry can in certain cases be significantly more efficient and robust. We have employed spherical neutron polarimetry to show that in \BaTi\ there are two possible magnetic spin structures. In combination with previously recorded neutron powder diffraction, we find that in \BaTi, the Cu spins are arranged in two-in-two-out manner with spins pointing approximately perpendicular to the CuO$_4$ motif. This spin structure is consistent with a strong DM interaction which naturally explains the large spin-gap observed in the magnetic spectrum of \BaTi\ \cite{kimura-natcomm-2016}. Given the rich physics in the $A$($B$O)Cu$_4$(PO$_4$)$_4$ family \cite{kimura-unpub}, our measurements should pave the way for future experimental and theoretical investigations of these intriguing materials.

\begin{acknowledgments}
The study was supported by the Swiss National Science Foundation and its Synergia network Mott Physics Beyond the Heisenberg Model (MPBH). K.K. is grateful for the funding received from the Japan Society for the Promotion of Science, Grants No. 16K05449 and a research grant from The Murata Science Foundation.
\end{acknowledgments}

\bibliography{shorttitles,biblio}

\begin{thebibliography}{20}%
\makeatletter
\providecommand \@ifxundefined [1]{%
 \@ifx{#1\undefined}
}%
\providecommand \@ifnum [1]{%
 \ifnum #1\expandafter \@firstoftwo
 \else \expandafter \@secondoftwo
 \fi
}%
\providecommand \@ifx [1]{%
 \ifx #1\expandafter \@firstoftwo
 \else \expandafter \@secondoftwo
 \fi
}%
\providecommand \natexlab [1]{#1}%
\providecommand \enquote  [1]{``#1''}%
\providecommand \bibnamefont  [1]{#1}%
\providecommand \bibfnamefont [1]{#1}%
\providecommand \citenamefont [1]{#1}%
\providecommand \href@noop [0]{\@secondoftwo}%
\providecommand \href [0]{\begingroup \@sanitize@url \@href}%
\providecommand \@href[1]{\@@startlink{#1}\@@href}%
\providecommand \@@href[1]{\endgroup#1\@@endlink}%
\providecommand \@sanitize@url [0]{\catcode `\\12\catcode `\$12\catcode
  `\&12\catcode `\#12\catcode `\^12\catcode `\_12\catcode `\%12\relax}%
\providecommand \@@startlink[1]{}%
\providecommand \@@endlink[0]{}%
\providecommand \url  [0]{\begingroup\@sanitize@url \@url }%
\providecommand \@url [1]{\endgroup\@href {#1}{\urlprefix }}%
\providecommand \urlprefix  [0]{URL }%
\providecommand \Eprint [0]{\href }%
\providecommand \doibase [0]{http://dx.doi.org/}%
\providecommand \selectlanguage [0]{\@gobble}%
\providecommand \bibinfo  [0]{\@secondoftwo}%
\providecommand \bibfield  [0]{\@secondoftwo}%
\providecommand \translation [1]{[#1]}%
\providecommand \BibitemOpen [0]{}%
\providecommand \bibitemStop [0]{}%
\providecommand \bibitemNoStop [0]{.\EOS\space}%
\providecommand \EOS [0]{\spacefactor3000\relax}%
\providecommand \BibitemShut  [1]{\csname bibitem#1\endcsname}%
\let\auto@bib@innerbib\@empty
\bibitem [{\citenamefont {Spaldin}\ \emph {et~al.}(2008)\citenamefont
  {Spaldin}, \citenamefont {Fiebig},\ and\ \citenamefont
  {Mostovoy}}]{spaldin-jpcm-2008}%
  \BibitemOpen
  \bibfield  {author} {\bibinfo {author} {\bibfnamefont {N.~A.}\ \bibnamefont
  {Spaldin}}, \bibinfo {author} {\bibfnamefont {M.}~\bibnamefont {Fiebig}}, \
  and\ \bibinfo {author} {\bibfnamefont {M.}~\bibnamefont {Mostovoy}},\ }\href
  {\doibase 10.1088/0953-8984/20/43/434203} {\bibfield  {journal} {\bibinfo
  {journal} {J. Phys.: Condens. Matter}\ }\textbf {\bibinfo {volume} {20}},\
  \bibinfo {pages} {434203} (\bibinfo {year} {2008})}\BibitemShut {NoStop}%
\bibitem [{\citenamefont {Kimura}\ \emph
  {et~al.}(2016{\natexlab{a}})\citenamefont {Kimura}, \citenamefont {Sera},\
  and\ \citenamefont {Kimura}}]{kimura-inorg-2016}%
  \BibitemOpen
  \bibfield  {author} {\bibinfo {author} {\bibfnamefont {K.}~\bibnamefont
  {Kimura}}, \bibinfo {author} {\bibfnamefont {M.}~\bibnamefont {Sera}}, \ and\
  \bibinfo {author} {\bibfnamefont {T.}~\bibnamefont {Kimura}},\ }\href
  {\doibase 10.1021/acs.inorgchem.5b02622} {\bibfield  {journal} {\bibinfo
  {journal} {Inorg. Chem.}\ }\textbf {\bibinfo {volume} {55}},\ \bibinfo
  {pages} {1002} (\bibinfo {year} {2016}{\natexlab{a}})}\BibitemShut {NoStop}%
\bibitem [{\citenamefont {Kato}\ \emph {et~al.}(2017)\citenamefont {Kato},
  \citenamefont {Kimura}, \citenamefont {Miyake}, \citenamefont {Tokunaga},
  \citenamefont {Matsuo}, \citenamefont {Kindo}, \citenamefont {Akaki},
  \citenamefont {Hagiwara}, \citenamefont {Sera}, \citenamefont {Kimura},\ and\
  \citenamefont {Motome}}]{kato-prl-2017}%
  \BibitemOpen
  \bibfield  {author} {\bibinfo {author} {\bibfnamefont {Y.}~\bibnamefont
  {Kato}}, \bibinfo {author} {\bibfnamefont {K.}~\bibnamefont {Kimura}},
  \bibinfo {author} {\bibfnamefont {A.}~\bibnamefont {Miyake}}, \bibinfo
  {author} {\bibfnamefont {M.}~\bibnamefont {Tokunaga}}, \bibinfo {author}
  {\bibfnamefont {A.}~\bibnamefont {Matsuo}}, \bibinfo {author} {\bibfnamefont
  {K.}~\bibnamefont {Kindo}}, \bibinfo {author} {\bibfnamefont
  {M.}~\bibnamefont {Akaki}}, \bibinfo {author} {\bibfnamefont
  {M.}~\bibnamefont {Hagiwara}}, \bibinfo {author} {\bibfnamefont
  {M.}~\bibnamefont {Sera}}, \bibinfo {author} {\bibfnamefont {T.}~\bibnamefont
  {Kimura}}, \ and\ \bibinfo {author} {\bibfnamefont {Y.}~\bibnamefont
  {Motome}},\ }\href {\doibase 10.1103/PhysRevLett.118.107601} {\bibfield
  {journal} {\bibinfo  {journal} {Phys. Rev. Lett.}\ }\textbf {\bibinfo
  {volume} {118}},\ \bibinfo {pages} {107601} (\bibinfo {year}
  {2017})}\BibitemShut {NoStop}%
\bibitem [{\citenamefont {Kimura}\ \emph
  {et~al.}(2016{\natexlab{b}})\citenamefont {Kimura}, \citenamefont
  {Babkevich}, \citenamefont {Sera}, \citenamefont {Toyoda}, \citenamefont
  {Yamauchi}, \citenamefont {Tucker}, \citenamefont {Martius}, \citenamefont
  {Fennell}, \citenamefont {Manuel}, \citenamefont {Khalyavin}, \citenamefont
  {Johnson}, \citenamefont {Nakano}, \citenamefont {Nozue}, \citenamefont
  {R{\o}nnow},\ and\ \citenamefont {Kimura}}]{kimura-natcomm-2016}%
  \BibitemOpen
  \bibfield  {author} {\bibinfo {author} {\bibfnamefont {K.}~\bibnamefont
  {Kimura}}, \bibinfo {author} {\bibfnamefont {P.}~\bibnamefont {Babkevich}},
  \bibinfo {author} {\bibfnamefont {M.}~\bibnamefont {Sera}}, \bibinfo {author}
  {\bibfnamefont {M.}~\bibnamefont {Toyoda}}, \bibinfo {author} {\bibfnamefont
  {K.}~\bibnamefont {Yamauchi}}, \bibinfo {author} {\bibfnamefont {G.~S.}\
  \bibnamefont {Tucker}}, \bibinfo {author} {\bibfnamefont {J.}~\bibnamefont
  {Martius}}, \bibinfo {author} {\bibfnamefont {T.}~\bibnamefont {Fennell}},
  \bibinfo {author} {\bibfnamefont {P.}~\bibnamefont {Manuel}}, \bibinfo
  {author} {\bibfnamefont {D.~D.}\ \bibnamefont {Khalyavin}}, \bibinfo {author}
  {\bibfnamefont {R.~D.}\ \bibnamefont {Johnson}}, \bibinfo {author}
  {\bibfnamefont {T.}~\bibnamefont {Nakano}}, \bibinfo {author} {\bibfnamefont
  {Y.}~\bibnamefont {Nozue}}, \bibinfo {author} {\bibfnamefont {H.~M.}\
  \bibnamefont {R{\o}nnow}}, \ and\ \bibinfo {author} {\bibfnamefont
  {T.}~\bibnamefont {Kimura}},\ }\href {\doibase 10.1038/ncomms13039}
  {\bibfield  {journal} {\bibinfo  {journal} {Nat. Commun.}\ }\textbf {\bibinfo
  {volume} {7}},\ \bibinfo {pages} {13039} (\bibinfo {year}
  {2016}{\natexlab{b}})}\BibitemShut {NoStop}%
\bibitem [{\citenamefont {Rodriguez-Carvajal}(1993)}]{fullprof}%
  \BibitemOpen
  \bibfield  {author} {\bibinfo {author} {\bibfnamefont {J.}~\bibnamefont
  {Rodriguez-Carvajal}},\ }\href {\doibase
  http://dx.doi.org/10.1016/0921-4526(93)90108-I} {\bibfield  {journal}
  {\bibinfo  {journal} {Physica B}\ }\textbf {\bibinfo {volume} {192}},\
  \bibinfo {pages} {55 } (\bibinfo {year} {1993})}\BibitemShut {NoStop}%
\bibitem [{Note1()}]{Note1}%
  \BibitemOpen
  \bibinfo {note} {The case of $\Gamma ^{(2)}_5$ can produce a magnetic
  structure with an amplitude modulated moment, which seems
  unlikely.}\BibitemShut {Stop}%
\bibitem [{\citenamefont {Walters}\ \emph {et~al.}(2009)\citenamefont
  {Walters}, \citenamefont {Perring}, \citenamefont {Caux}, \citenamefont
  {Savici}, \citenamefont {Gu}, \citenamefont {Lee}, \citenamefont {Ku},\ and\
  \citenamefont {Zaliznyak}}]{walters-natphys-2009}%
  \BibitemOpen
  \bibfield  {author} {\bibinfo {author} {\bibfnamefont {A.~C.}\ \bibnamefont
  {Walters}}, \bibinfo {author} {\bibfnamefont {T.~G.}\ \bibnamefont
  {Perring}}, \bibinfo {author} {\bibfnamefont {J.-S.}\ \bibnamefont {Caux}},
  \bibinfo {author} {\bibfnamefont {A.~T.}\ \bibnamefont {Savici}}, \bibinfo
  {author} {\bibfnamefont {G.~D.}\ \bibnamefont {Gu}}, \bibinfo {author}
  {\bibfnamefont {C.-C.}\ \bibnamefont {Lee}}, \bibinfo {author} {\bibfnamefont
  {W.}~\bibnamefont {Ku}}, \ and\ \bibinfo {author} {\bibfnamefont {I.~A.}\
  \bibnamefont {Zaliznyak}},\ }\href {\doibase 10.1038/nphys1405} {\bibfield
  {journal} {\bibinfo  {journal} {Nature Phys.}\ }\textbf {\bibinfo {volume}
  {5}},\ \bibinfo {pages} {867} (\bibinfo {year} {2009})}\BibitemShut {NoStop}%
\bibitem [{\citenamefont {Chatterji}(2006)}]{chatterji-book}%
  \BibitemOpen
  \bibfield  {author} {\bibinfo {author} {\bibfnamefont {T.}~\bibnamefont
  {Chatterji}},\ }\href@noop {} {\emph {\bibinfo {title} {Neutron scattering
  from magnetic materials}}}\ (\bibinfo  {publisher} {Elsevier},\ \bibinfo
  {address} {Amsterdam; London},\ \bibinfo {year} {2006})\BibitemShut {NoStop}%
\bibitem [{\citenamefont {Squires}(2012)}]{squires-book}%
  \BibitemOpen
  \bibfield  {author} {\bibinfo {author} {\bibfnamefont {G.~L.}\ \bibnamefont
  {Squires}},\ }\href {\doibase 10.1017/CBO9781139107808} {\emph {\bibinfo
  {title} {Introduction to the Theory of Thermal Neutron Scattering}}}\
  (\bibinfo  {publisher} {Cambridge University Press},\ \bibinfo {year}
  {2012})\BibitemShut {NoStop}%
\bibitem [{\citenamefont {Shirane}\ \emph {et~al.}(2002)\citenamefont
  {Shirane}, \citenamefont {Shapiro},\ and\ \citenamefont
  {Tranquada}}]{shirane-book}%
  \BibitemOpen
  \bibfield  {author} {\bibinfo {author} {\bibfnamefont {G.}~\bibnamefont
  {Shirane}}, \bibinfo {author} {\bibfnamefont {S.~M.}\ \bibnamefont
  {Shapiro}}, \ and\ \bibinfo {author} {\bibfnamefont {J.~M.}\ \bibnamefont
  {Tranquada}},\ }\href@noop {} {\emph {\bibinfo {title} {Neutron scattering
  with a triple-axis spectrometer: basic techniques}}}\ (\bibinfo  {publisher}
  {Cambridge University Press},\ \bibinfo {address} {New York},\ \bibinfo
  {year} {2002})\BibitemShut {NoStop}%
\bibitem [{\citenamefont {Roessli}\ and\ \citenamefont
  {B{\"o}ni}(2001)}]{roessli-book-2001}%
  \BibitemOpen
  \bibfield  {author} {\bibinfo {author} {\bibfnamefont {B.}~\bibnamefont
  {Roessli}}\ and\ \bibinfo {author} {\bibfnamefont {P.}~\bibnamefont
  {B{\"o}ni}}\ }(\bibinfo  {publisher} {Academic Press},\ \bibinfo {address}
  {London},\ \bibinfo {year} {2001})\ p.\ \bibinfo {pages} {1242}\BibitemShut
  {NoStop}%
\bibitem [{\citenamefont {Blume}(1963)}]{blume-pr-1963}%
  \BibitemOpen
  \bibfield  {author} {\bibinfo {author} {\bibfnamefont {M.}~\bibnamefont
  {Blume}},\ }\href {\doibase 10.1103/PhysRev.130.1670} {\bibfield  {journal}
  {\bibinfo  {journal} {Phys. Rev.}\ }\textbf {\bibinfo {volume} {130}},\
  \bibinfo {pages} {1670} (\bibinfo {year} {1963})}\BibitemShut {NoStop}%
\bibitem [{\citenamefont {Izyumov}\ and\ \citenamefont
  {Maleyev}(1962)}]{izyumov-sovphys-1962}%
  \BibitemOpen
  \bibfield  {author} {\bibinfo {author} {\bibfnamefont {Y.}~\bibnamefont
  {Izyumov}}\ and\ \bibinfo {author} {\bibfnamefont {S.}~\bibnamefont
  {Maleyev}},\ }\href@noop {} {\bibfield  {journal} {\bibinfo  {journal} {Sov.
  Phys. JETP}\ }\textbf {\bibinfo {volume} {14}},\ \bibinfo {pages} {1168}
  (\bibinfo {year} {1962})}\BibitemShut {NoStop}%
\bibitem [{\citenamefont {Fischer}(1997)}]{fischer-physicab-1997}%
  \BibitemOpen
  \bibfield  {author} {\bibinfo {author} {\bibfnamefont {W.~E.}\ \bibnamefont
  {Fischer}},\ }\href {\doibase 10.1016/S0921-4526(97)00260-3} {\bibfield
  {journal} {\bibinfo  {journal} {Physica B}\ }\textbf {\bibinfo {volume} {{\bf
  234}}},\ \bibinfo {pages} {1202} (\bibinfo {year} {1997})}\BibitemShut
  {NoStop}%
\bibitem [{\citenamefont {Semadeni}\ \emph {et~al.}(2001)\citenamefont
  {Semadeni}, \citenamefont {Roessli},\ and\ \citenamefont
  {B\"{o}ni}}]{semadeni-physicab-2001}%
  \BibitemOpen
  \bibfield  {author} {\bibinfo {author} {\bibfnamefont {F.}~\bibnamefont
  {Semadeni}}, \bibinfo {author} {\bibfnamefont {B.}~\bibnamefont {Roessli}}, \
  and\ \bibinfo {author} {\bibfnamefont {P.}~\bibnamefont {B\"{o}ni}},\ }\href
  {\doibase 10.1016/S0921-4526(00)00860-7} {\bibfield  {journal} {\bibinfo
  {journal} {Physica B}\ }\textbf {\bibinfo {volume} {{\bf 297}}},\ \bibinfo
  {pages} {152} (\bibinfo {year} {2001})}\BibitemShut {NoStop}%
\bibitem [{\citenamefont {Janoschek}\ \emph {et~al.}(2007)\citenamefont
  {Janoschek}, \citenamefont {Klimko}, \citenamefont {G\"{a}hler},
  \citenamefont {Roessli},\ and\ \citenamefont
  {B\"{o}ni}}]{janoschek-physicab-2007}%
  \BibitemOpen
  \bibfield  {author} {\bibinfo {author} {\bibfnamefont {M.}~\bibnamefont
  {Janoschek}}, \bibinfo {author} {\bibfnamefont {S.}~\bibnamefont {Klimko}},
  \bibinfo {author} {\bibfnamefont {R.}~\bibnamefont {G\"{a}hler}}, \bibinfo
  {author} {\bibfnamefont {B.}~\bibnamefont {Roessli}}, \ and\ \bibinfo
  {author} {\bibfnamefont {P.}~\bibnamefont {B\"{o}ni}},\ }\href {\doibase
  10.1016/j.physb.2007.02.074} {\bibfield  {journal} {\bibinfo  {journal}
  {Physica B}\ }\textbf {\bibinfo {volume} {{\bf 397}}},\ \bibinfo {pages}
  {125} (\bibinfo {year} {2007})}\BibitemShut {NoStop}%
\bibitem [{Note2()}]{Note2}%
  \BibitemOpen
  \bibinfo {note} {We note that the $\chi ^2_\nu $ only includes the random
  errors and not the systematic ones, resulting in values somewhat larger than
  1.}\BibitemShut {Stop}%
\bibitem [{\citenamefont {Kimura}\ \emph {et~al.}()\citenamefont {Kimura},
  \citenamefont {Toyoda}, \citenamefont {Babkevich}, \citenamefont {Yamauchi},
  \citenamefont {Sera}, \citenamefont {Nassif}, , \citenamefont {R{\o}nnow},\
  and\ \citenamefont {Kimura}}]{kimura-unpub}%
  \BibitemOpen
  \bibfield  {author} {\bibinfo {author} {\bibfnamefont {K.}~\bibnamefont
  {Kimura}}, \bibinfo {author} {\bibfnamefont {M.}~\bibnamefont {Toyoda}},
  \bibinfo {author} {\bibfnamefont {P.}~\bibnamefont {Babkevich}}, \bibinfo
  {author} {\bibfnamefont {K.}~\bibnamefont {Yamauchi}}, \bibinfo {author}
  {\bibfnamefont {M.}~\bibnamefont {Sera}}, \bibinfo {author} {\bibfnamefont
  {V.}~\bibnamefont {Nassif}}, , \bibinfo {author} {\bibfnamefont {H.~M.}\
  \bibnamefont {R{\o}nnow}}, \ and\ \bibinfo {author} {\bibfnamefont
  {T.}~\bibnamefont {Kimura}},\ }\href@noop {} {\bibinfo  {journal}
  {unpublished}\ }\BibitemShut {NoStop}%
\bibitem [{\citenamefont {Dzyaloshinsky}(1958)}]{dzyaloshinsky-jopcs-1958}%
  \BibitemOpen
\bibfield  {journal} {  }\bibfield  {author} {\bibinfo {author} {\bibfnamefont
  {I.}~\bibnamefont {Dzyaloshinsky}},\ }\href {\doibase
  https://doi.org/10.1016/0022-3697(58)90076-3} {\bibfield  {journal} {\bibinfo
   {journal} {J. Phys. Chem. Solids}\ }\textbf {\bibinfo {volume} {4}},\
  \bibinfo {pages} {241 } (\bibinfo {year} {1958})}\BibitemShut {NoStop}%
\bibitem [{\citenamefont {Moriya}(1960)}]{moriya-pr-1960}%
  \BibitemOpen
  \bibfield  {author} {\bibinfo {author} {\bibfnamefont {T.}~\bibnamefont
  {Moriya}},\ }\href {\doibase 10.1103/PhysRev.120.91} {\bibfield  {journal}
  {\bibinfo  {journal} {Phys. Rev.}\ }\textbf {\bibinfo {volume} {120}},\
  \bibinfo {pages} {91} (\bibinfo {year} {1960})}\BibitemShut {NoStop}%
\end{thebibliography}%

\end{document}